\documentclass[twocolumn,showpacs,preprintnumbers,amsmath,amssymb]{revtex4}
\input epsf
\usepackage{graphicx}% Include figure files
\usepackage{dcolumn}% Align table columns on decimal point
\usepackage{bm}% bold math

\begin{document}

\preprint{cond-mat.other}

\title{The effect of pressure and spin on the isotopic composition of
  ferrous iron dissolved in 
periclase and silicate perovskite} % Force line breaks with \\

\author{James R. Rustad, Piotr Zarzycki, Maryali P. Sauceda, and Qing-zhu Yin}
\affiliation{Department of Geology, University of California-Davis,
 One Shields Avenue, Davis, CA 95616 U.~S.~A.}
  \email{jrrustad@ucdavis.edu}

\date{\today}

\begin{abstract}
We perform density functional calculations of the equilibrium 
$^{57}$Fe/$^{54}$Fe ratios for ferrous iron dissolved in periclase 
and MgSiO$_3$ perovskite at the pressures and temperatures of the 
Earth's mantle.  Pressure increases the partitioning of $^{57}$Fe 
into both phases by a factor of three from the Earth's surface 
to the core-mantle boundary.  In ferropericlase, a large contribution 
to this increase comes from the electronic transition from 
high-spin to low-spin iron.  Our calculations demonstrate that 
pressure-induced fractionation can play a major role in determining planetary 
iron-isotope composition.
\end{abstract}

% PACS, the Physics and Astronomy
                             % Classification Scheme.
%\keywords{Suggested keywords}%Use showkeys class option if keyword
                              %display desired
\maketitle

%\begin{figure}
%\includegraphics{fig_1.eps}% Here is how to import EPS art
%\caption{\label{fig:epsart} A figure caption. The figure captions are
%automatically numbered.}
%\end{figure}
 
The distribution of iron-isotopes between and within planetary bodies
provides constraints on their histories of accretion and differentiation
 \cite{beard-2004,weyer-2005,poitrasson-2007}.  A significant fraction
 of the iron in the silicate Earth 
is dissolved in 
periclase (Fe$_x$Mg$_{1-x}$O) ({\bf fep}) and silicate perovskite 
(Fe$_x$Mg$_{1-x}$)SiO$_3$ ({\bf fepv}) which make up the bulk of
the Earth's lower mantle \cite{kesson-1998}.  Uptake 
by {\bf fep} and {\bf fepv}  has almost certainly  
influenced the distribution of iron-isotopes 
in the Earth, but little is known definitively about the equilibrium iron 
isotope compositions of these phases at lower-mantle pressures and
temperatures.  
Empirical estimates, based on the vibrational
density of states for the iron-sublattice obtained from M\" ossbauer
spectroscopy and inelastic nuclear resonant X-ray
scattering (INRXS) techniques, are starting to address this issue \cite{polyakov-2007,polyakov-2008}.   
 
Equilibrium isotopic signatures should become 
heavier with increasing pressure as bonds are compressed and
vibrational frequencies increase.   Because the ionic
radius of low-spin iron is smaller than high-spin iron, iron-isotope signatures also
may be made heavier by the spin transition
for Fe$^{2+}$ \cite{fyfe-1960,cohen-1997}, which, in {\bf fep}, is predicted to occur within 
the Earth's mantle \cite{speziale-2005,tsuchiya-2006}.   
The reduced partition function ratio
($\beta$), can be used to make predictions of the effect of 
pressure and electronic state on the possible isotopic fractionations 
associated with {\bf fep} and {\bf fepv}.
In the harmonic approximation $\beta$ is defined by \cite{bigeleisen-1947}:
 \begin{equation}
\beta = \left( {{Q_h}\over{Q_l}} \right)=\Pi_i %{{{u_h}_i}\over{{u_l}_i}} 
{{e^{-{{u_h}_i}/2}}\over{1-e^{-{u_h}_i}}}{{1-e^{-{u_l}_i}}\over{e^{{-{u_l}_i}/2}}}
\end{equation}
where ${u_{(h,l)}}_i = \hbar c 2\pi{\omega_{(h,l)}}_i/kT$, $h$ and $l$ refer 
to the heavy isotope and light isotope, respectively (here $^{57}$Fe
and $^{54}$Fe), $T$ is the temperature, and the product 
runs over all frequencies $\omega_i$.  The equilibrium constant for isotopic 
exchange $\alpha$ between phases $i$ and $j$ is given 
as $\alpha_{i,j}=\beta_i/\beta_j$; $\alpha_{i,j} > 1$ implies 
that the heavy isotope is concentrated in phase $i$.
Here we use density functional electronic structure methods to
calculate $\beta$ for Fe$^{2+}$ 
in {\bf fepv} and {\bf fep}
at pressures and temperatures near that of the core-mantle boundary.
We have found that $\beta$ increases dramatically with pressure
and is strongly affected by the high-spin to low-spin transition
in {\bf fep}.
 
Density functional calculations
have  made important contributions to understanding the properties of lower 
mantle phases
\cite{cohen-1997,tsuchiya-2006, stackhouse-2007}.  They also
have been used to estimate isotope 
fractionation factors in iron-bearing phases such as 
$\rm{Fe}^{2+,3+}$(aq) {\it via}  
harmonic vibrational frequencies \cite{hill-2008,anbar-2005}.  
Here, we use  these
methods on clusters representative of the coordination environment 
of Fe$^{2+}$ in {\bf fep} and {\bf fepv}.  The clusters can be treated with  
standard all-electron quantum chemistry methods, avoiding potential 
problems of the iron pseudopotential and making a stronger connection 
with existing studies of vibrational frequencies and the
spin-crossover transition in the quantum 
chemistry literature, which can guide the choice of 
exchange-correlation functionals and basis set \cite{zein-2007}.  Such methods 
allow affordable use of hybrid exchange-correlation functionals which can 
make accurate estimates of the spin-crossover energies
in extensive studies on molecular systems \cite{fouqueau-2004} and
have provided the best estimates of isotope fractionation factors
\cite{rustad-2008a}.  
The hybrid functionals complement calculations using the local-density and
generalized-gradient approximations \cite{cohen-1997,stackhouse-2007}
and studies using the Hubbard model
\cite{tsuchiya-2006}.

Iron atoms are assumed to vibrate
in coordination environments characteristic of {\bf fep} and 
{\bf fepv}, similar to the site-specific
vibrational density of states approach adopted in
Ref.~\cite{polyakov-2008} .  
The models consist of a free ``core'' 
surrounded by a rigid shell of oxygen atoms fixed in their ideal lattice
positions.  The vibrational modes of the core
within the cavity that displace the iron center 
are assumed to approximate the vibrational modes
that govern equilibrium isotopic fractionation in the crystal.
The cores of the clusters are shown in Figure 1.  The {\bf fepv} site
consists of a single iron atom and eight coordinating oxide atoms.  
All silicon and magnesium atoms attached to these eight oxygens 
are also included in the core cluster.  The core is surrounded 
by a layer of oxygen atoms completing the coordination shells of 
the silicon and magnesium atoms.  The other bonds into the outer 
layer of oxygen atoms coming from ``external'' magnesium and silicon
atoms, are represented by link atoms having a charge 
equal to the Pauling bond strength (the charge divided by the 
coordination number) contributed into the oxygen atom.  The link atoms are placed along
the broken bonds one \aa ngstrom  away from the oxygen atom.  This procedure 
ensures a neutral, autocompensated cluster.  The {\bf fep} cluster, centered 
on the octahedral iron site, is constructed 
in the same manner.  This simple technique accurately reproduces
 calculations of isotope fractionation in 
carbonates and other oxide minerals with periodic boundary conditions \cite{rustad-2008a,rustad-2008b}. 

%\begin{figure}
%\center
%\includegraphics[scale=0.10]{prl_fig1a.pdf}% Here is how to import EPS art
%\end{figure}
\begin{figure}
\center
\includegraphics[scale=0.17]{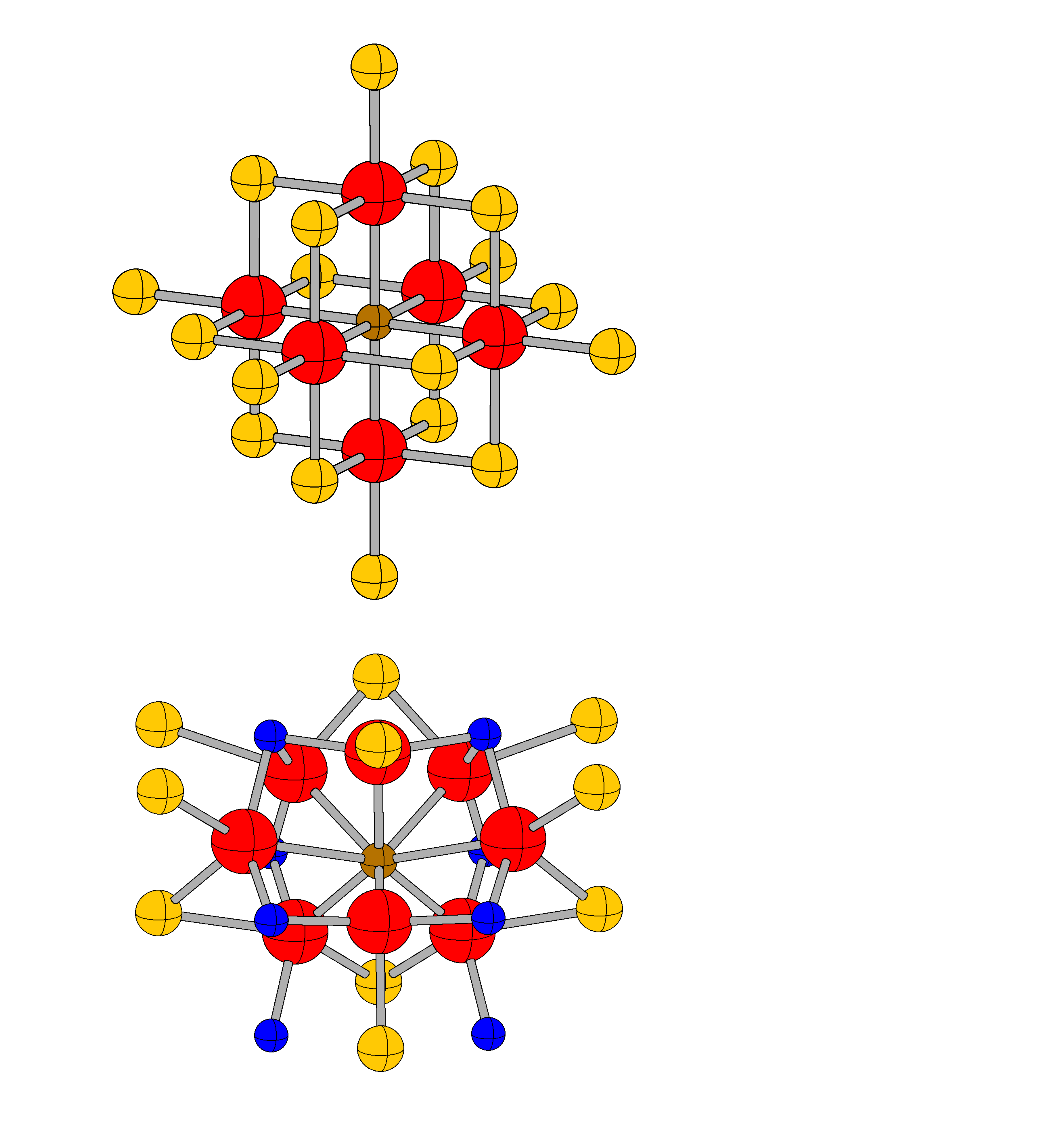}
\caption{\label{fig1}  Core atoms of molecular clusters  
used to model fe-periclase ({\bf fep}, above) and fe-perovskite ({\bf
  fepv}, below).  
The iron atom (brown) is centered within  its coordination environment of oxygen (red), 
magnesium (yellow) and silicon (blue).  These core atoms
are fully relaxed within a shell of oxygen atoms, fixed in their ideal
lattice positions.  This rigid outer shell
of oxygen atoms is capped by hydrogen atoms having fractional charges as described
in the text.}
\end{figure}
    
The electronic structure calculations, carried out with the PQS quantum 
chemistry package of Pulay, Baker, and
co-workers (http://www.pqs-chem.com).  The rigid outer shell uses the
3-21G basis on both the oxygen 
atoms and the link atoms (which use hydrogen basis functions).  The optimized 
atoms use the 6-31G* basis for O, Mg, and Si, 
and the $m$6-31G* basis for Fe \cite{mitin-2003}.  
   The B3LYP 
exchange-correlation 
functional, as implemented in the PQS code,
was used for the exchange-correlation functional.   For {\bf fepv} outer-shell oxygen positions
were taken from Ref.~\cite{vanpeteghem-2006} at 0 GPa and from
electronic structure calculations \cite{iitaka-2004} at 120 GPa.
Lattice parameters for {\bf fep}
were obtained from a recent universal P-V-T equation of state for periclase
\cite{garai-2008}.  The high-pressure lattice parameters were taken at
120 GPa and at the highest available temperature of 3000~K, even 
though we evaluate Eq.~1 at 4000~K.
The low-pressure lattice parameters were taken at
0 GPa and 300 K.   

For the {\bf fepv}
cluster, we enforce mirror symmetry through the center
of the cluster perpendicular to the {\bf c} crystallographic
axes, giving the cluster point symmetry $C_s$.  
For the low-spin {\bf fep} cluster we retain the $T_h$ symmetry group;
this is relaxed in the high-spin cluster which was run with no
symmetry constraints to accomodate possible Jahn-Teller 
distortions.  The positions of the 
central iron atom, its coordinating oxygen atoms, and the 
magnesium and silicon atoms attached to the coordinating 
oxygen atoms are optimized within the rigid outer shell 
of the cluster (oxygen atoms and link atoms).  Relaxed bond lengths 
in the primary coordination 
sphere of the iron for all clusters are given in Table 1.  

\begin{table}
\caption{\label{tab:table1}Calculated bond lengths, in \AA , with
multiplicities given in parentheses, if different from unity.}
\begin{ruledtabular}
\begin{tabular}{cccccc}
 \multicolumn{2} {c} {\bf fepv}& \multicolumn{4} {c} {\bf fep}\\
 \multicolumn{2} {c} {high-spin} \ & \multicolumn{2} {c} {high-spin} & \multicolumn{2} {c} {low-spin}\\
 \cline{1-2}\cline{3-4}\cline{5-6}\\[-7 pt]
 0 & 120 & 0 & 120 & 0 & 120\\
\hline\\[-7 pt]
1.958 &1.857     & 2.213 (2) & 1.978 (2)  & 2.213 (6)  & 1.959 (6) \\
 2.149&1.927     & 2.276 (4) & 2.008 (4)       &        &       \\
2.034 (2)&1.866 (2) &    &        &        &       \\
2.462 (2)&2.062 (2) &    &        &        &       \\
 2.478 (2)&2.212 (2) &   &        &        &       \\
\end{tabular}
\end{ruledtabular}
\end{table}

The Hessian matrix for the free atoms was calculated analytically 
by projecting out the degrees of freedom of the fixed atoms. 
 Harmonic vibrational frequencies were calculated for the core atoms 
with both $^{54}$Fe and $^{57}$Fe, and used to estimate 
$\beta$ in the harmonic approximation according to Eq. 1.
The $\beta$ defined by Eq.~1 was divided by the high temperature limiting
value $\beta_{HT}$ according to standard convention
\cite{chacko-1991}.

Calculations on low-spin Fe$^{2+}$ in the {\bf fepv} cage in $C_s$ symmetry
gave a vibrational mode with an imaginary frequency perpendicular to the 
mirror plane.
low-spin Fe$^{2+}$ would therefore ``rattle'' unfavorably in a cage with $C_s$
symmetry, qualitatively consistent
with the idea the Fe$^{2+}$ is not likely to undergo a pressure-induced 
high-spin to low-spin transition
in {\bf fepv} in the Earth's mantle \cite{hofmeister-2006}.  We did
not consider low-spin Fe$^{2+}$ further in the {\bf fepv} cluster.  
Fe$^{3+}$ in the octahedral Si$^{4+}$ site may undergo spin crossover
within the mantle \cite{stackhouse-2007}, but this is not considered here.  

The results of the calculations are shown in Table 2, 
given as 10$^3$$\ln(\beta /\beta_{HT})$, for 
temperatures of 300 K and 4000 K, the latter temperature 
being close to that of the
core-mantle boundary.  First, pressures at the core mantle 
boundary increase the $\beta$ 
factors for both phases by a factor of 2.5-3 
relative to the Earth's surface.  
For {\bf fep} at pressures near the core-mantle boundary, half of the enrichment 
can be attributed to the high-spin to low-spin transition in {\bf fep} 
A much smaller difference is predicted at ambient pressures.  
Although we do not claim to be able to quantitatively model 
the energetics of the spin transition here,
we remark that the electronic energy for the cluster with 
low-spin iron {\bf fep} 
is approximately 1.5 eV above that for the cluster with 
high-spin iron at 0 GPa.     
In the high-pressure cluster, however, the electronic
energy of the low-spin cluster is about 0.4 eV {\em below} 
that for high-spin cluster.
The $\beta$ values for {\bf fep} given in Table 2 are in reasonable 
agreement with those reported in \cite{polyakov-2008}.  The 
high-pressure value of $\beta$ is 
is somewhat lower than our calculations.  Part of the reason 
for this is that the high-pressure
value of $\beta$ based on INRXS measurements is given at 
109 GPa whereas the calculations
were carried out at $\approx$ 120 GPa.

\begin{table}
\caption{\label{tab:table2}Reduced partition function ratios
  10$^3$ln($\beta /\beta_{HT}$), calculated for $^{57}$Fe amd
  $^{54}$Fe, according to Eq. 1 and as
  described in the text.}
\begin{ruledtabular}
\begin{tabular}{ccccccc}
 & \multicolumn{2} {c} {\bf fepv}& \multicolumn{4} {c} {\bf fep}\\
 & \multicolumn{2} {c} {high-spin} \ & \multicolumn{2} {c} {high-spin} & \multicolumn{2} {c} {low-spin}\\
 \cline{2-3}\cline{4-5}\cline{6-7}\\[-7 pt]
 P (GPa)& 0 & 120 & 0 & 120 & 0 & 120\\
\hline\\[-7 pt]
300 K & 9.86 & 24.2 & 7.04 (7.8)\footnotemark[1] & 16.5  & 8.1  & 24.1 (20.1)\footnotemark[1]\footnotemark[2] \\
4000 K\footnotemark[3]& 0.06 & 0.15 & 0.04 &  0.10 & 0.05 & 0.15 \\
\end{tabular}
\end{ruledtabular}
\footnotetext[1]{taken from Ref.~\onlinecite{polyakov-2008}}
\footnotetext[2]{P = 109 GPa}
\footnotetext[3]{the periclase lattice parameters are taken at 120~GPa 
and 3000~K from Ref.~\onlinecite{garai-2008}, but the $\beta$
  was evaluated at 4000~K in Eq.~1}
\end{table}

The contributions to $\beta$ for the high-spin and low-spin {\bf fep}
clusters are shown in Fig.~2.  Contributions to $\beta$ are spread out
among seven main fractionating modes.  Characteristic motions
associated with these modes
are given in Fig.~3.  For both the high-spin and low-spin
configurations, the largest contribution to $\beta$ comes from the
fundamental vibration of the central iron against the cage of
surrounding oxygens and magnesium ions at 308 (302) cm$^{-1}$ for
$^{54}$Fe($^{57}$Fe) in low-spin
{\bf fep}.  This is an accoustic mode
with the coordinating oxygen atoms and magnesium atoms moving 
with the iron, but with less displacement.  Although this mode
makes the biggest contribution to $\beta$, the difference between
the high-spin and low-spin contributions is not very large.  The next most important
fractionating mode is at 491.3 (489.80) cm$^{-1}$, an optical
mode involving the iron atom vibrating against its own foxygen
coordination shell and their attached magnesium atoms.  In the high-spin {\bf fep}
cluster, this mode occurs at somewhat lower frequencies, and is 
spread out among several similar modes.  This mode does not fractionate
iron as intensely in the high-spin cluster as it does in the low-spin cluster; $\beta$ is nearly four
per mil lower than the low-spin cluster after passing through
these vibrational modes.  The mode at 560.3 (559.8) cm$^{-1}$ in the
low-spin cluster, mostly involving motion of the equatorial FeO$_4$
plane against the oxygens at the apex, also makes a
much larger contribution to $\beta$ than the corresponding vibration
in the high-spin cluster.  The next two vibrational modes make
roughly equal contributions to $\beta$ in both the high- and low-spin
clusters.  The last two modes, involving primary stretching 
 of the FeO$_6$ octahedron, with the iron in phase with the magnesium atoms, again make
stronger contributions for the low-spin cluster than for the
high-spin cluster.

\begin{figure}
\center
\includegraphics[scale=0.3]{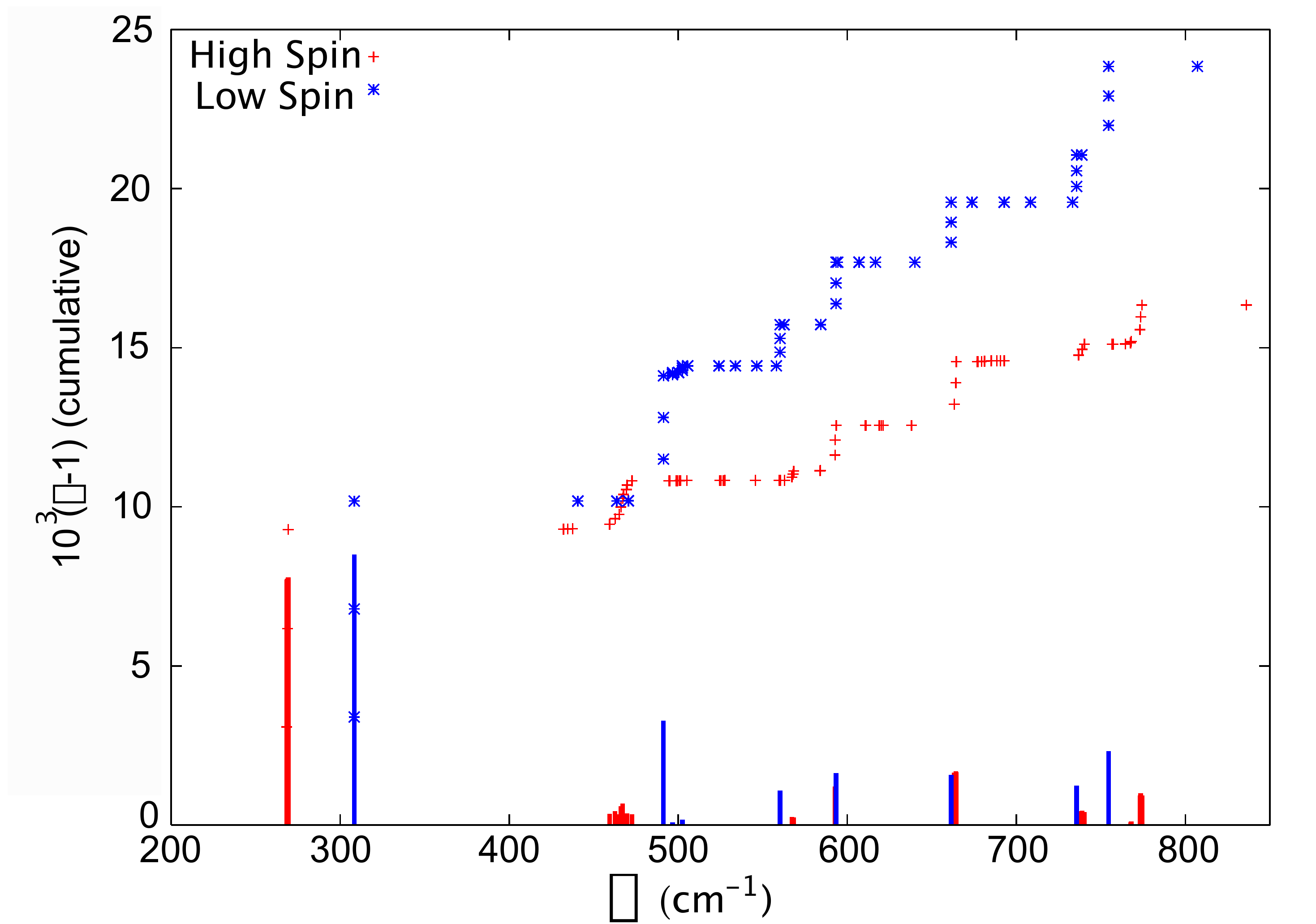}
\caption{\label{fig2}  Cumulative value of ($\beta$-1) for high-spin
  and low-spin {\bf fep} cluster at P$\approx$ 120 GPa.  Contributions of different
  vibrational modes to $\beta$ are given by impulses.}
\end{figure}

\begin{figure}
\center
\includegraphics[scale=0.35]{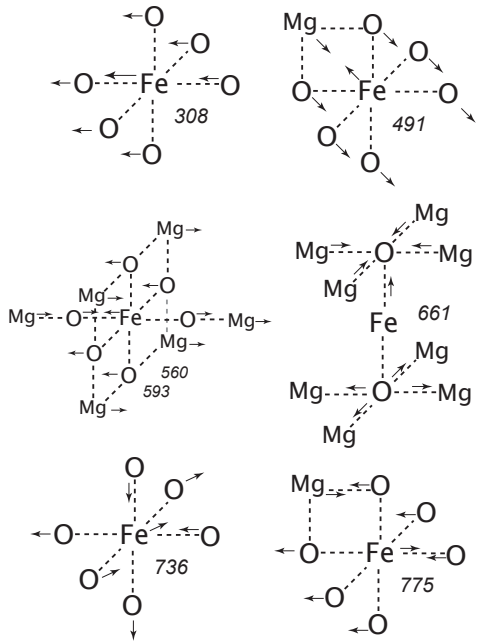}
\caption{\label{fig3} Major fractionating vibrational modes shown in
  Fig.~2.  Numbers in italics give vibrational frequencies in
  cm$^{-1}$.  The modes at 560 cm$^{-1}$ and 593 cm$^{-1}$ involve
similar displacements of the central iron atom.}
\end{figure}

The magnitudes of the increases in $\beta$ with pressure has
implications
 for the isotopic signatures of 
accreting planetary bodies.
First, an individual planet in isotopic equilibrium 
should be isotopically stratified 
with heavier isotopes progressively concentrated at depth within 
the planet.  Such a stratification
could possibly be established during a ``magma ocean'' stage as iron 
isotopes sample coordination environments
in liquid silicates that would be similar to {\bf fep} 
or {\bf fepv}.  An interesting question is whether a spin transition
would take place in the melt.  If coordination environments
are structurally closer to {\bf fepv} \cite{stixrude-2005}, the spin transition may be
supressed and the isotopic signature could be lighter than a solid of
the same composition.  Freezing of a magma ocean might be
expected to give rise to an isotopically stratified mantle 
with the heavy isotopes accumulating preferentially
at the base, depending on the relative densities of the liquid and
solid phases.  

If an equilibrium isotopic profile, with $^{57}$Fe/$^{54}$Fe increasing with
depth, can be established after impact
events, evaporative loss of surface material
during extreme heating would be isotopically light, thus 
the larger planets should be isotopically heavier than the 
smaller planets, because the larger 
planets can shield greater reserviors of material at high-pressure 
where the $\beta$ factor is
larger.  This may be part of the reason why the Earth-Moon 
system appears to be isotopically heavier
than Mars and Vesta \cite{poitrasson-2004}.

The pressure dependence of $\beta$ 
is sensitive to crystallographic environment; $\beta_{{\rm\bf fepv}}$
increases more than a factor of two over high-spin $\beta_{{\rm\bf
    fep}}$, probably because of the greater compressibility
of the coordination shell in {\bf fepv}.  However, the increase
is compensated completely at the core-mantle 
boundary by the spin transition in {\bf fep}.
The similarity between $\beta_{{\rm\bf fepv}}$ and $\beta_{{\rm\bf
    fep}}$ indicates 
 little differential isotopic
fractionation between these phases after the
spin-transition takes place in the lower mantle.  Above this
transition, we expect {\bf fepv} to be enriched in $^{57}$Fe over {\bf fep}
by a factor of two.

\begin{acknowledgments}
This work was funded by the U.~S. Department of Energy, division of Basic Energy
Sciences, grant DE-FG02-04ER15498, and NSF grant EAR-08-14242.
\end{acknowledgments}

\bibliography{rustad}% Produces the bibliography via BibTeX.

\end{document}